\documentclass[epjCONF]{svjour}
\usepackage{graphics}
\usepackage[varg]{txfonts} 
\usepackage[latin1]{inputenc}

\newcommand{\pt}{$p_\mathrm{T}$}

\session-title{Resonance Workshop at UT Austin}
\begin{document}
\title{Measurements of the electron-positron continuum in ALICE}
\author{Christoph Baumann\thanks{\email{cbaumann@ikf.uni-frankfurt.de}} for the ALICE collaboration}
\institute{Institut f\"ur Kernphysik, Goethe-Universit\"at Frankfurt}
\abstract{
The status of the analysis of electron-positron pairs measured by ALICE in pp collisions at $\sqrt{s} = 7$~TeV and central Pb-Pb collisions at $\sqrt{s_\mathrm{NN}}=2.76$~TeV is presented. Key questions and the main challenges of the analysis are discussed on the basis of first raw invariant mass spectra for both collision systems.
} 
\maketitle

A hot and dense state of matter, the quark-gluon plasma, can be created in ultrarelativistic heavy-ion collisions.

Electromagnetic probes are unique probes of the time evolution of the QGP, as they carry information from all stages of the collision.
Earlier experiments have shown that the measurement of electron-positron and muon pairs offers access to many interesting observables:
The spectra of vector resonances ($\omega$, $\phi$, J/$\Psi$) and the pronounced contribution, at LHC energies, from open charm decays. At lower center-of-mass energies, an excess over the yield
expected from the vacuum decay of hadrons has been observed at low momenta for masses below  $m_\omega$~\cite{Agakishiev:1997au,Arnaldi:2006jq,Arnaldi:2008fw,Adare:2009qk,Ruan:2012za}. This enhancement can be explained a thermal production of electron-positron pairs and an in-medium modification of the $\rho$~\cite{Rapp:2009yu,Linnyk:2012pu}.

Additionally, it is possible to measure the production of direct photons via electron-positron pairs at small invariant masses, which originate from virtual photons. Using this method, the PHENIX experiment has observed an excess photon production in Au-Au central collisions over the expected value from pp collisions that may be of thermal origin~\cite{Adare:2008ab}.

The ALICE detector at the CERN-LHC is uniquely suited for the measurement of low-mass electron-positron pairs. A full description of the experiment
is given in~\cite{Aamodt:2008zz}.
Employing different detector technologies, the experiment provides particle identification and tracking for momenta down to $100$~MeV/$c$, which is a unique feature among the LHC experiments.
For the analysis presented here, we combine information from the ITS, TPC and TOF detectors to select electron (positron) candidates. These detectors are located at mid-rapidity ($|y| < 0.9$) and provide full azimuthal acceptance.

Results from two data sets are presented in these proceedings: pp collisions at $\sqrt{s} = 7$~TeV and Pb-Pb collisions at $\sqrt{s_\mathrm{NN}} = 2.76$~TeV

\section{Analysis of the pp Data}

\begin{figure}
\begin{center}
\resizebox{0.45\columnwidth}{!}{
  \includegraphics{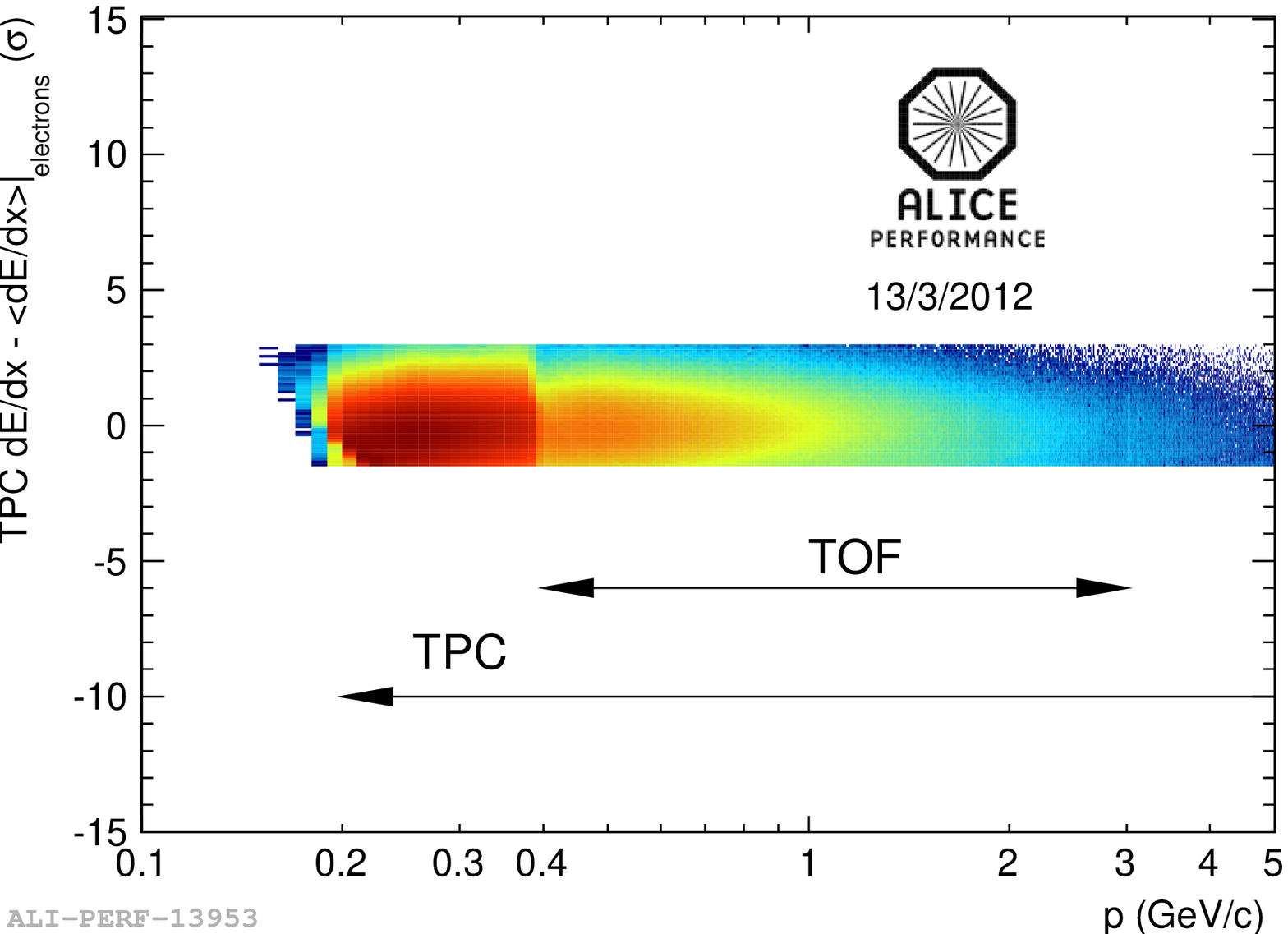} }
\resizebox{0.45\columnwidth}{!}{
  \includegraphics{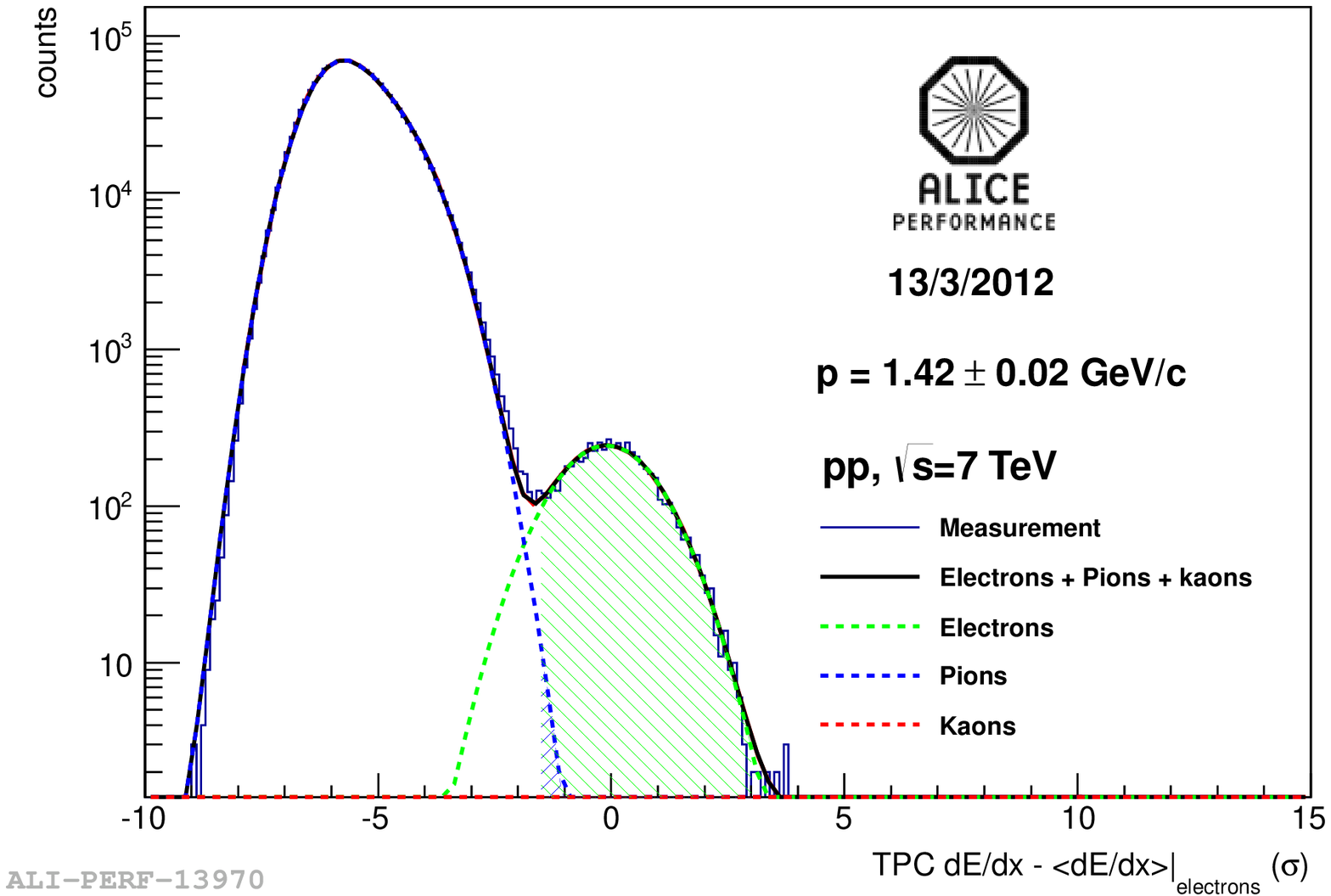} }
\end{center}
\caption{The left figure shows the distribution of the TPC dEdx-signal, which is given in units standard deviations w.r.t. to the expected mean dEdx-signal of electron, as a function of the track momentum after the application of all track quality and particle identification (PID) cuts. The right panel shows the projection of the TPC signal before the application of the PID cuts for an exemplary momentum interval. The part of the distribution selected for analysis is shown a green shaded area.}
\label{fig:dEdxpp}       
\end{figure}

For the pp data, 300M minimum-bias events have been available for analysis. Figure~\ref{fig:dEdxpp} shows the TPC signal distribution for the tracks selected for the analysis. Besides passing basic quality criteria, a TPC signal for each track within $-1.5 < \sigma_{e} < 3$ of the expected mean electron signal for a given momentum was required.

For tracks with $p_\mathrm{T} > 0.4$~GeV/$c$, the tracks are additionally constrained, by requiring a TOF signal within $3\sigma$ of the expected electron signal at the corresponding momentum. As shown in the right panel of Figure~\ref{fig:dEdxpp}, this leads to a negligible contamination from other particles, most of all charged pions, over the full momentum range used for this analysis.

\begin{figure}
\begin{center}
\resizebox{0.9\columnwidth}{!}{
  \includegraphics{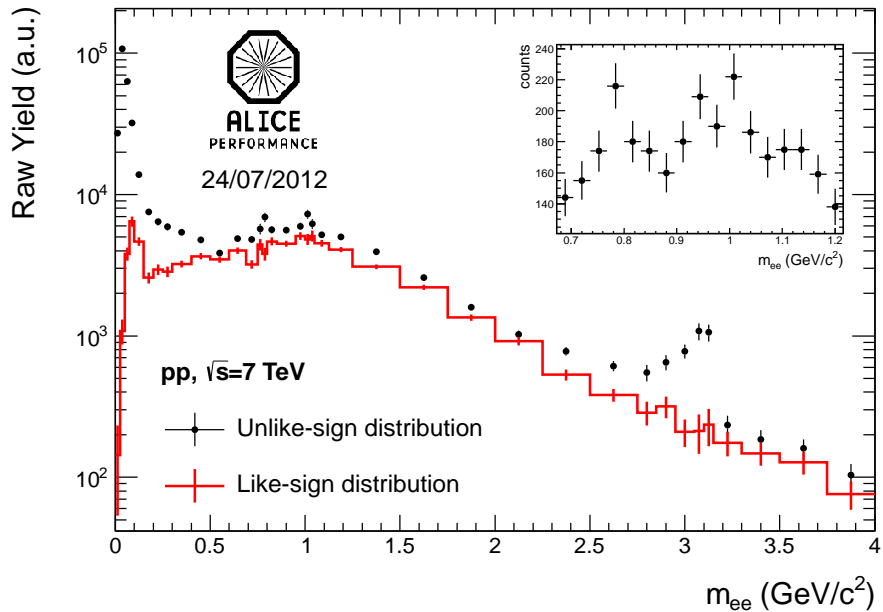} }
\end{center}
\caption{Electron-positron unlike-sign and like-sign invariant mass distributions integrated over \pt~measured in pp collisions at $\sqrt{s} = 7$~TeV. The like-sign distribution has been corrected to account for the difference in acceptance between like-sign and unlike-sign pairs.  }
\label{fig:ULLSpp}       
\end{figure}

To determine he invariant mass distribution of the dielectron signal all opposite-sign pair combinations of electron and positron candidates are determined in each event. The combinatorial background is calculated from same-event like-sign pairs: $Y_\mathrm{comb.BG.} = 2 \sqrt{Y_{++} Y_{--} } \cdot R$, where \\$R = Y_{+-,mixed} / (2 \sqrt{Y_{++,mixed} Y_{--,mixed} } )$ accounts for the acceptance difference between like-sign and unlike-sign pairs, which is modeled using a mixed-events technique (Figure~\ref{fig:ULLSpp}).

\begin{figure}
\begin{center}
\resizebox{0.75\columnwidth}{!}{
  \includegraphics{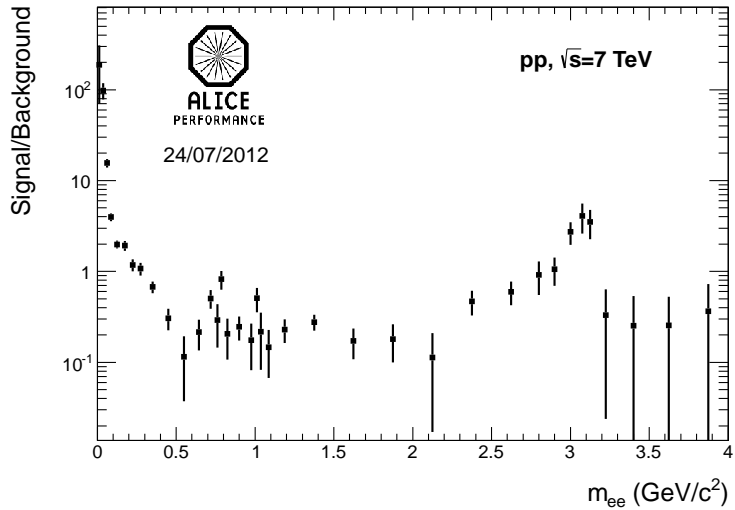} }
\end{center}
\caption{Signal to background ratio integrated over \pt~in pp collisions.}
\label{fig:SBpp}       
\end{figure}

\begin{figure}
\begin{center}
\resizebox{0.75\columnwidth}{!}{
  \includegraphics{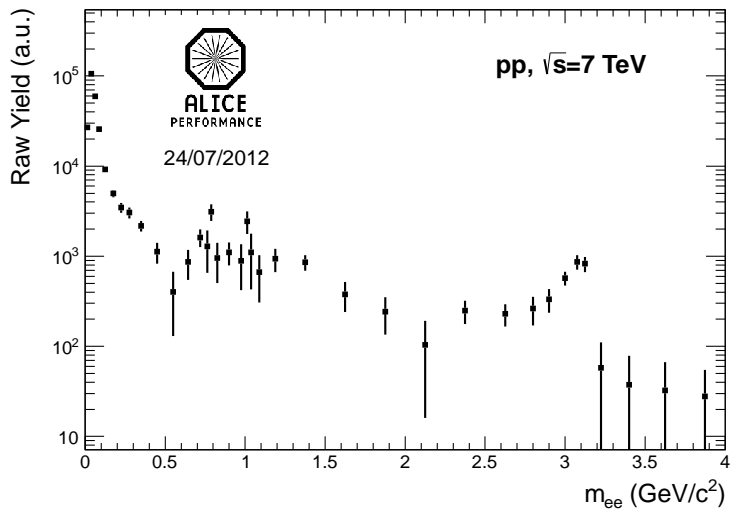} }
\end{center}
\caption{Background subtracted dielectron signal integrated over \pt~in pp collisions.}
\label{fig:Signalpp}       
\end{figure}

To reduce the combinatorial background, pairs with opening angles $\theta < 0.035$~rad are removed from the sample, as such pairs are predominantly originating from photon conversions.
The signal-to-background ratio for this measurement varies between $10^{-1} - 1$ as shown in Figure~\ref{fig:SBpp}.

This value is comparable to results obtained by ALICE in the dimuon channel at forward rapidities. They are complementary to the dielectron measurements, details are given in~\cite{ALICE:2011ad}.

Subtracting the like-sign distribution from the unlike-sign yield gives the raw electron-positron yield shown in Figure~\ref{fig:Signalpp}.
Within the statistical uncertainties, the yield is significant up to the J/$\Psi$ mass, the peaks of the $\omega$ and $\phi$ mesons are clearly distinguishable. The $p_\mathrm{T}$ dependence of the resonance yields will be compared in the future to the results obtained in the $\mu$ and $K$ decay channels.

Currently, the acceptance and efficiency corrections are being determined and a hadronic cocktail calculation that also accounts for the contribution from open-charm decays is under development. With these, it will be possible to compare the spectrum to theoretical expectations assuming vacuum decay of the hadrons and to measure or set limits on the charm cross-section.

\section{Analysis of the Pb-Pb Data}
In the 2011 beamtime ALICE has recorded enhanced samples of central and semi-central events in addition to minimum-bias data for Pb-Pb collisions at $\sqrt{s_\mathrm{NN}} = 2.76$~TeV.
In total, 8M minimum bias, 27M central and 32M semi-central events are available for analysis.
While the general analysis method is identical to the approach used for pp collisions, stricter cuts for the track selection have been chosen to account for the increase of the combinatorial background due to the higher event multiplicities: a signal within $3\sigma$ of the expected electron signal in both TPC and TOF for $p_\mathrm{T} > 0.4$~GeV/$c$ is required for electron candidates.

A main focus of this analysis is the search for a thermal contribution to the photon spectrum, measuring internal conversions in the electron-positron channel. The kinematical region relevant for such pairs is characterized by $p_{\mathrm{T},ee} > 1.0$~GeV/$c$ and $m_{ee} < 0.5$~GeV/$c^2$.
The virtual photon yield from internal conversions can be gauged by fitting the measured distribution with a function of the form $f_{total} (m_{ee})  = r \cdot f_{hadronic} (m_{ee}) + (1-r) f_{photon} (m_{ee})$, where $f_{hadronic}$ describes the shape of the hadronic contributions to the spectrum and $f_{photon}$ the shape of the photonic contributions, where both are independently normalized. $r$ can then be interpreted as the fraction of photons coming from internal conversions in the inclusive photon spectrum. Such a fit is typically made at masses around 200-300~MeV/$c^2$, where the spectrum is no longer dominated by the contribution from $\pi^0$-Dalitz decays.

\begin{figure}[hbt]
\resizebox{0.5\columnwidth}{!}{
  \includegraphics{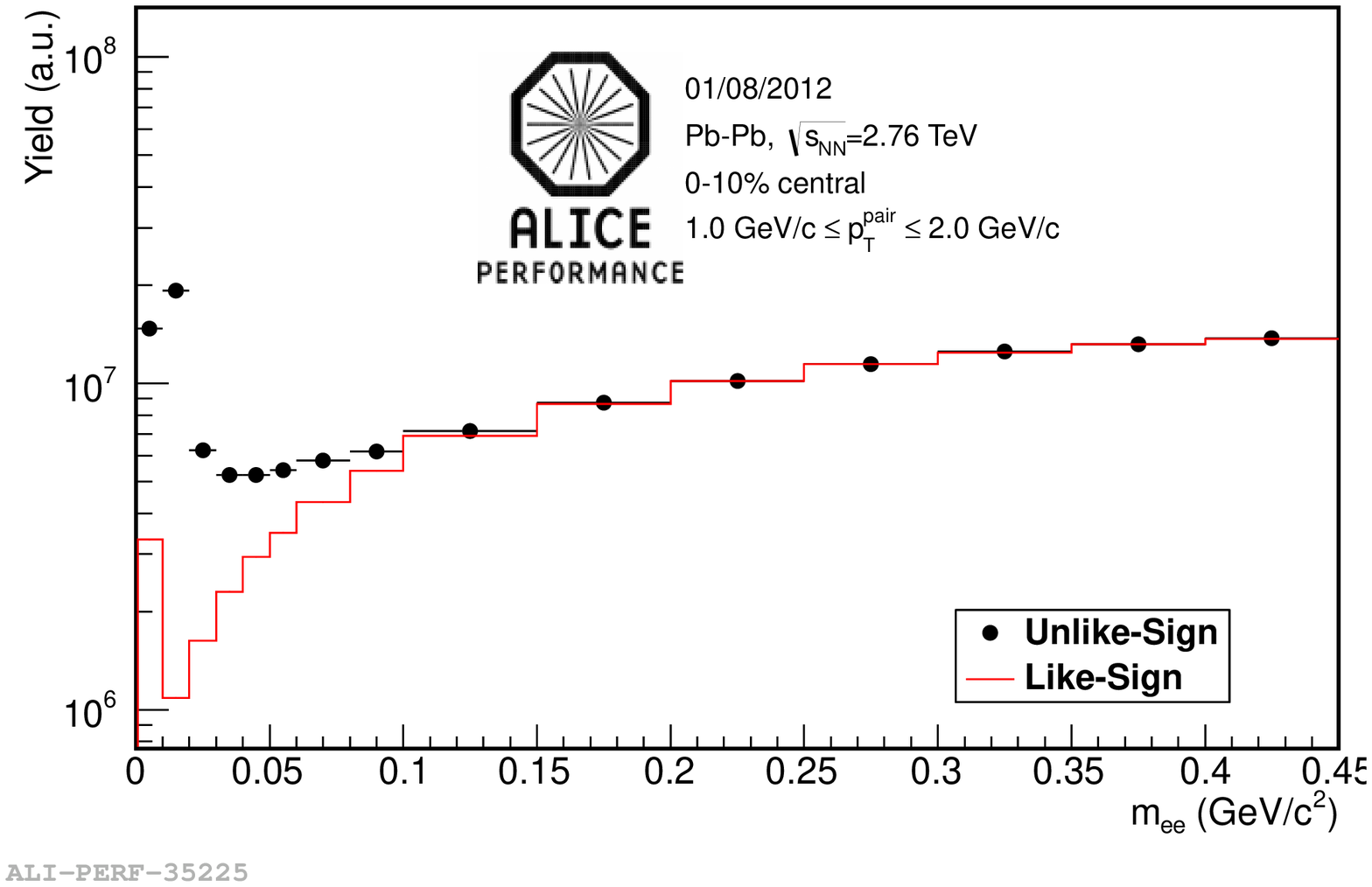} }
\resizebox{0.5\columnwidth}{!}{
  \includegraphics{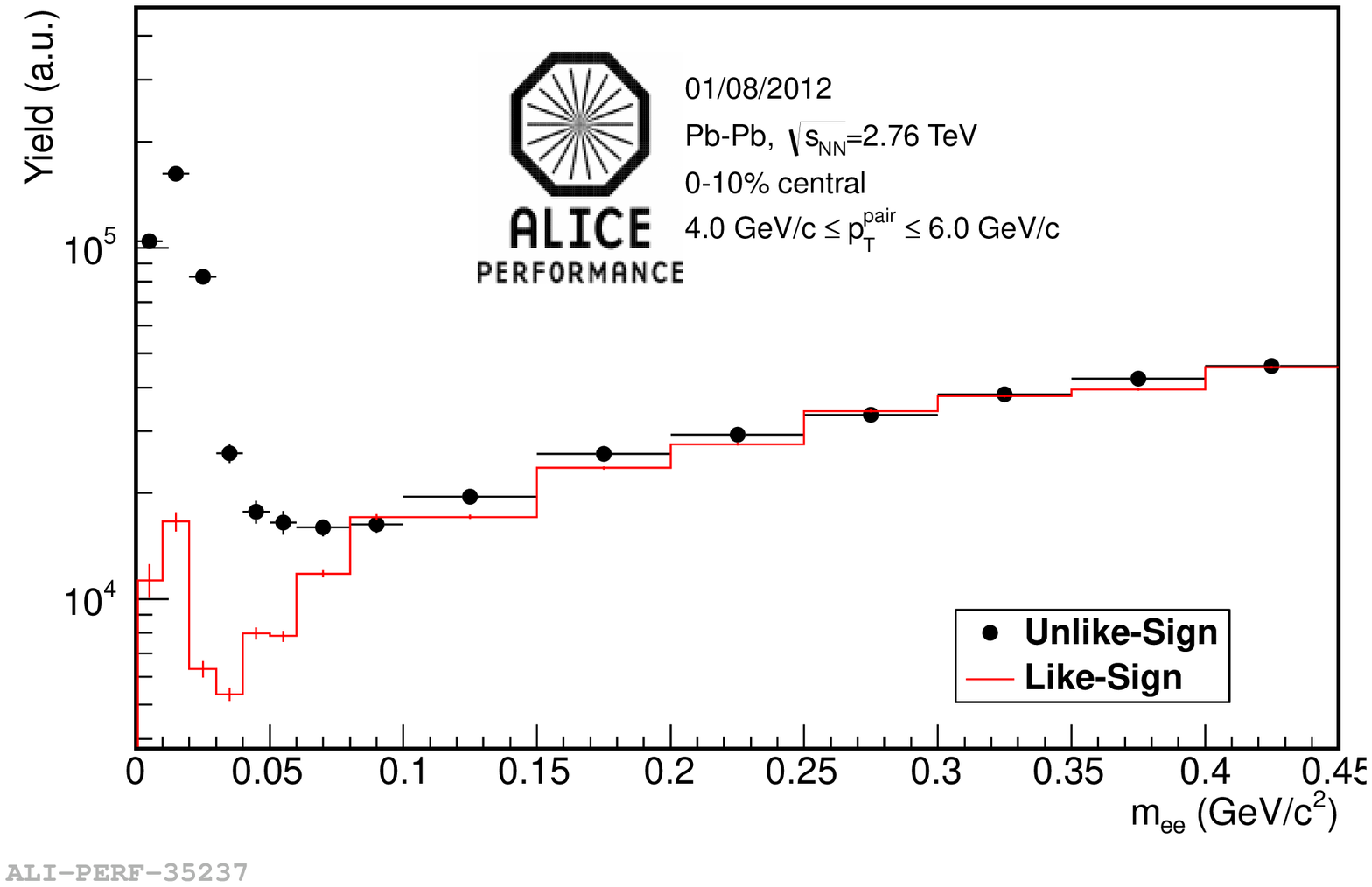} }
\caption{Unlike-Sign and same-event like-sign distributions for two \pt~intervals in central Pb-Pb collisions.}
\label{fig:ULLSPbPbCentral}       
\end{figure}

\begin{figure}[hbt]
\resizebox{0.5\columnwidth}{!}{
  \includegraphics{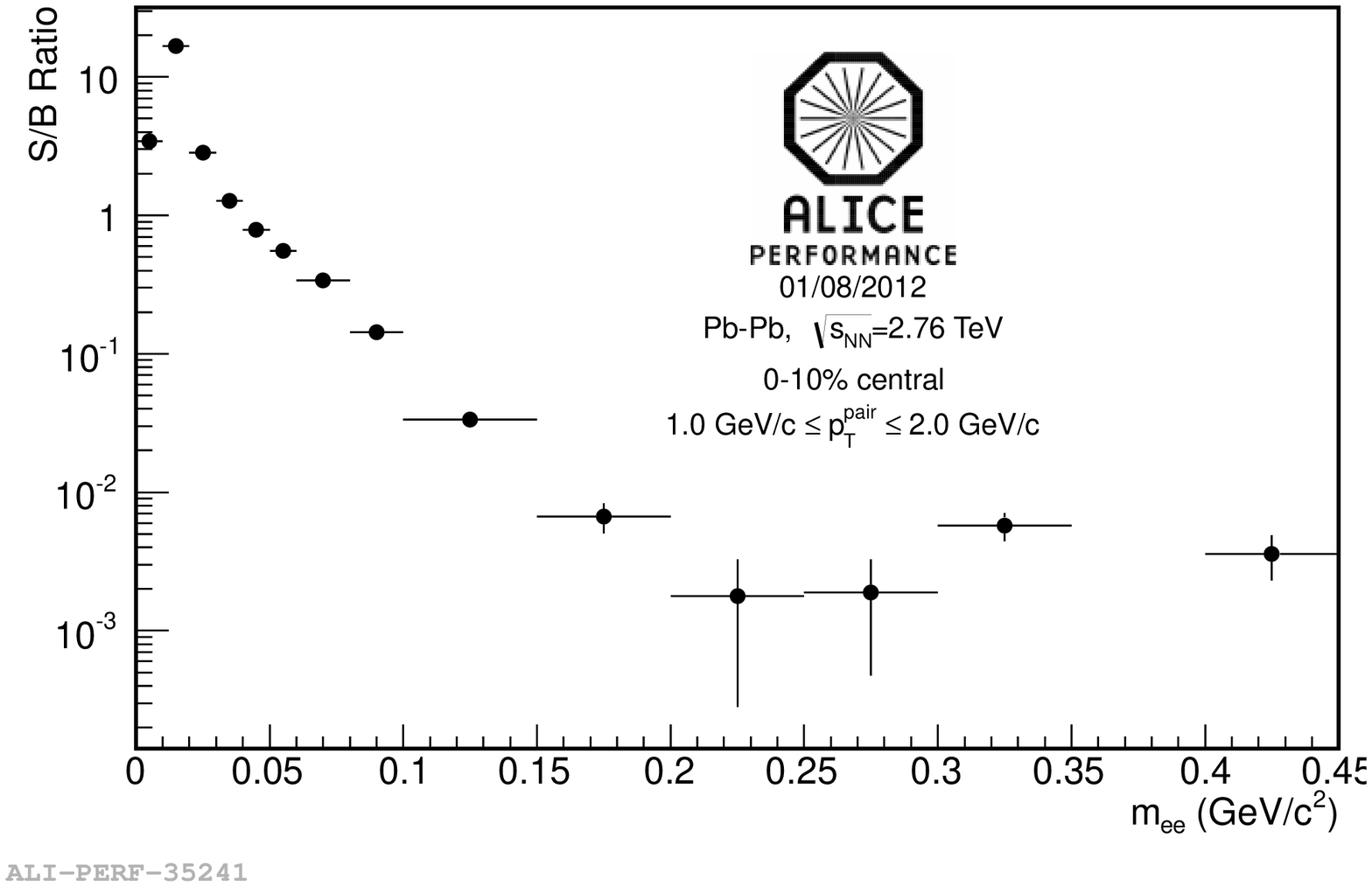} }
\resizebox{0.5\columnwidth}{!}{
  \includegraphics{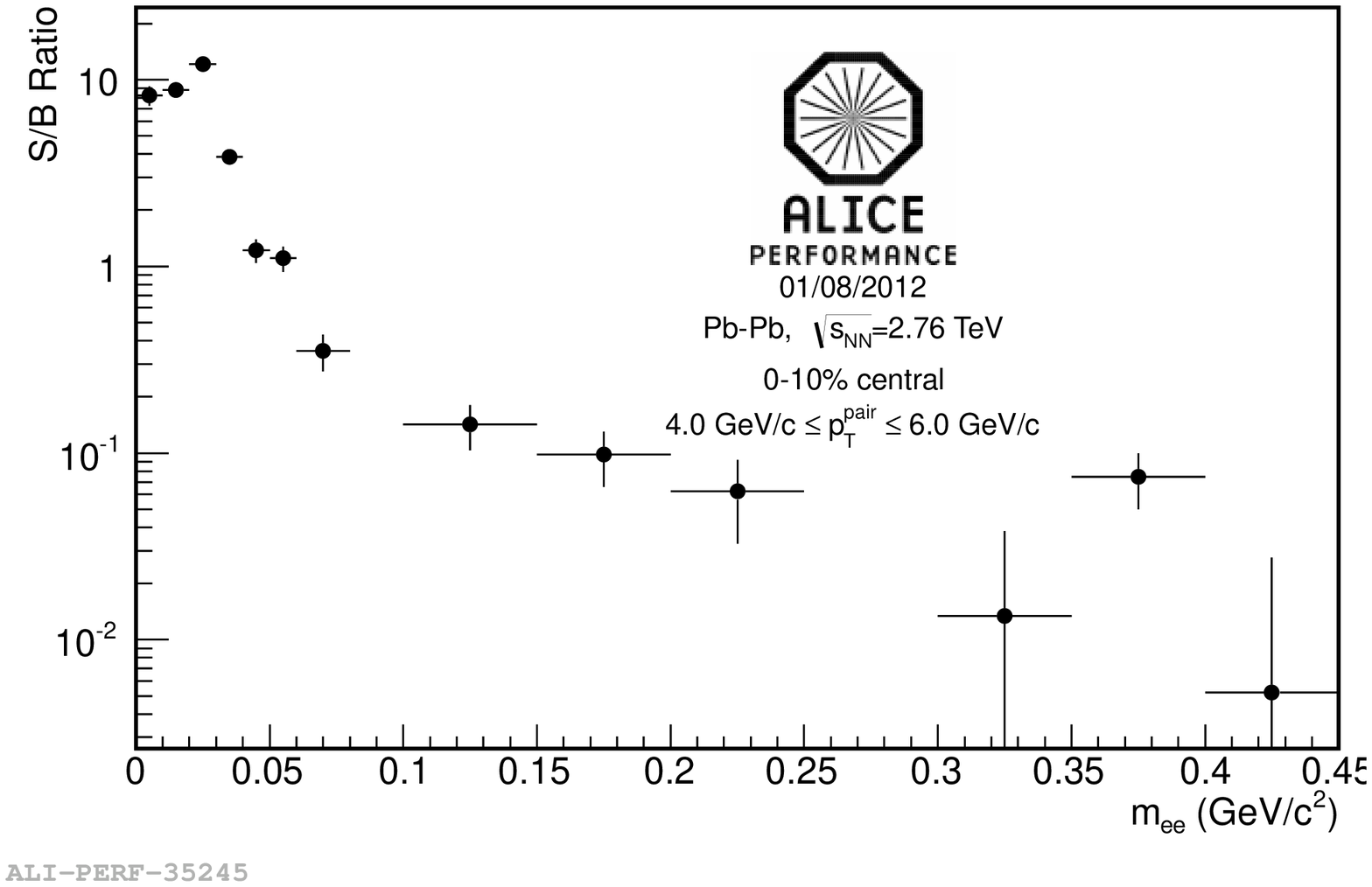} }
\caption{Signal to background ratio for two \pt~intervals in central Pb-Pb collisions.}
\label{fig:SBPbPbCentral}       
\end{figure}

The raw yield of such pairs was extracted in several $p_{\mathrm{T},ee}$ intervals following the method described above. Figure~\ref{fig:ULLSPbPbCentral} shows the unlike-sign and like-sign distributions for two $p_\mathrm{T}$ intervals as an example. The signal-to-background ratio shows, as can be expected, a dependence on the $p_\mathrm{T}$ region. A typical value for  $m_{ee} \approx 300$~MeV/$c^2$ is $S/B = 10^{-2}$ (Figure~\ref{fig:SBPbPbCentral}).

\begin{figure}
\resizebox{0.9\columnwidth}{!}{
  \includegraphics{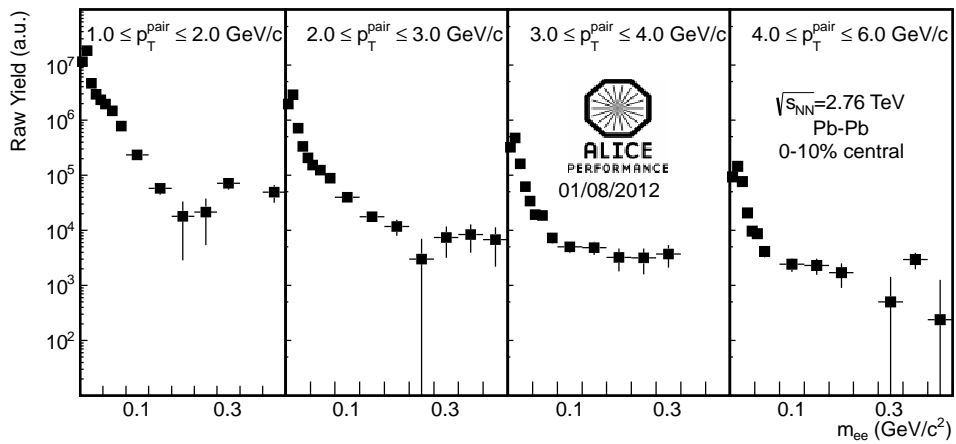} }
\caption{Background subtracted dielectron signal for several \pt~intervals in central Pb-Pb collisions.}
\label{fig:SignalPbPb}       
\end{figure}

The background subtracted dielectron spectrum is shown in Figure~\ref{fig:SignalPbPb}. At low invariant masses electron-positron pair from conversions of real photons contribute significantly to the spectrum. Different approaches to identify and reject these with good efficiency are currently under evaluation.

To allow quantitative conclusions on the dielectron production, these raw spectra need to be corrected for acceptance and efficiency. Furthermore, a hadronic cocktail calculation including contributions from open charm and bottom decays is being developed.

\section{Summary}
We have presented first results on the measurement of electron-positron pairs in both pp and Pb-Pb collisons measured with the ALICE experiment. Making use of its unique PID capabilities at low momenta, a raw electron-positron pair yield has been extracted in both collision systems. Using the pp results as a base-line, the search for a possible thermal contribution to the electron-positron spectrum at low invariant masses in the Pb-Pb data is a main focus of the ongoing analysis.

\end{document}